\documentclass[aps,prl,floatfix,twocolumn,superscriptaddress]{revtex4}

\usepackage{latexsym}
\usepackage{graphicx}
\usepackage{rotating}
\usepackage{hyperref}
\usepackage{amsmath,amssymb,amsfonts}
\usepackage{bm}
\usepackage{color}

\newcommand{\dx}{d$_{x2-y2}$ }
\newcommand{\psig}{p$_\sigma$ }

\def\onlinecite#1{\cite{#1}}

\newcommand{\up}{\uparrow}
\newcommand{\dn}{\downarrow}

\newcommand{\vk}{\bold{k}}

\begin{document}

\title{Strength of Correlations in electron and hole doped cuprates}
\author{C\'edric Weber}
\author{Kristjan Haule}
\author{Gabriel Kotliar}
\affiliation{Department of Physics, Rutgers University,  Piscataway, NJ 08854, USA}


\maketitle

\textbf{High temperature superconductivity was achieved by introducing
  holes in a parent compound consisting of copper oxide layers
  separated by spacer layers. Realizations of this phenomena has been
  achieved in multiple crystal structures and has been the subject of
  numerous investigations and extensive controversy. 
  In a small number
  of copper oxide based materials, it is possible to dope the parent
  compound with electrons
  \cite{armitage_review,tokura_electron_doped,takagi}
  , and their physical properties are bearing some
  similarities but also significant differences from the hole doped
  counterparts. For example, in the electron doped materials the
  antiferromagnetic phase is much more robust than the superconducting
  phase, while the normal state has a resistivity with a quadratic
  temperature dependence which is expected in normal metals rather
  than the linear temperature dependence observed in the hole doped
  systems. Here, we use a recently developed first principles method,
  to study the electron doped cuprates and elucidate the deep physical
  reasons why their behavior is so different than the hole doped
  materials.  The crystal structure of the electron doped materials,
  characterized by a lack of oxygen in the apical position, results in
  a parent compound which is a Slater insulator. Namely, a material
  where the insulating behavior is the result of the presence of
  magnetic long range order. This is in sharp contrast with the hole
  doped materials, where the parent compound is a Mott charge transfer
  insulator, namely a material which is insulating due to the strong
  electronic correlations but not the magnetic order.  We study the
  evolution of the angle resolved photoemission spectra and the
  optical properties of the normal state of the electron doped
  cuprates as a function of doping, clarifying how their unique
  position close to, but below the metal to charge transfer insulator
  transition, accounts for their surprising differences from the hole
  doped cuprates.}

We study the prototypical electron doped compound Nd$_{2-x}$Ce$_x$CuO$_4$ (NCCO).
This is a single-layer material and therefore we compare it to
La$_{2-x}$Sr$_x$CuO4 (LSCO), a single layer hole doped material in the
related $T$ structure.
We use a realistic theoretical approach, e.g. the Local Density
Approximation combined with the Dynamical Mean Field Theory
(LDA+DMFT)~\cite{OLD_GABIS_REVIEW,lda_dmft_held}, and focus on intermediate energy
quantities.
\begin{figure}
\begin{center}
\includegraphics[width=0.9\columnwidth]{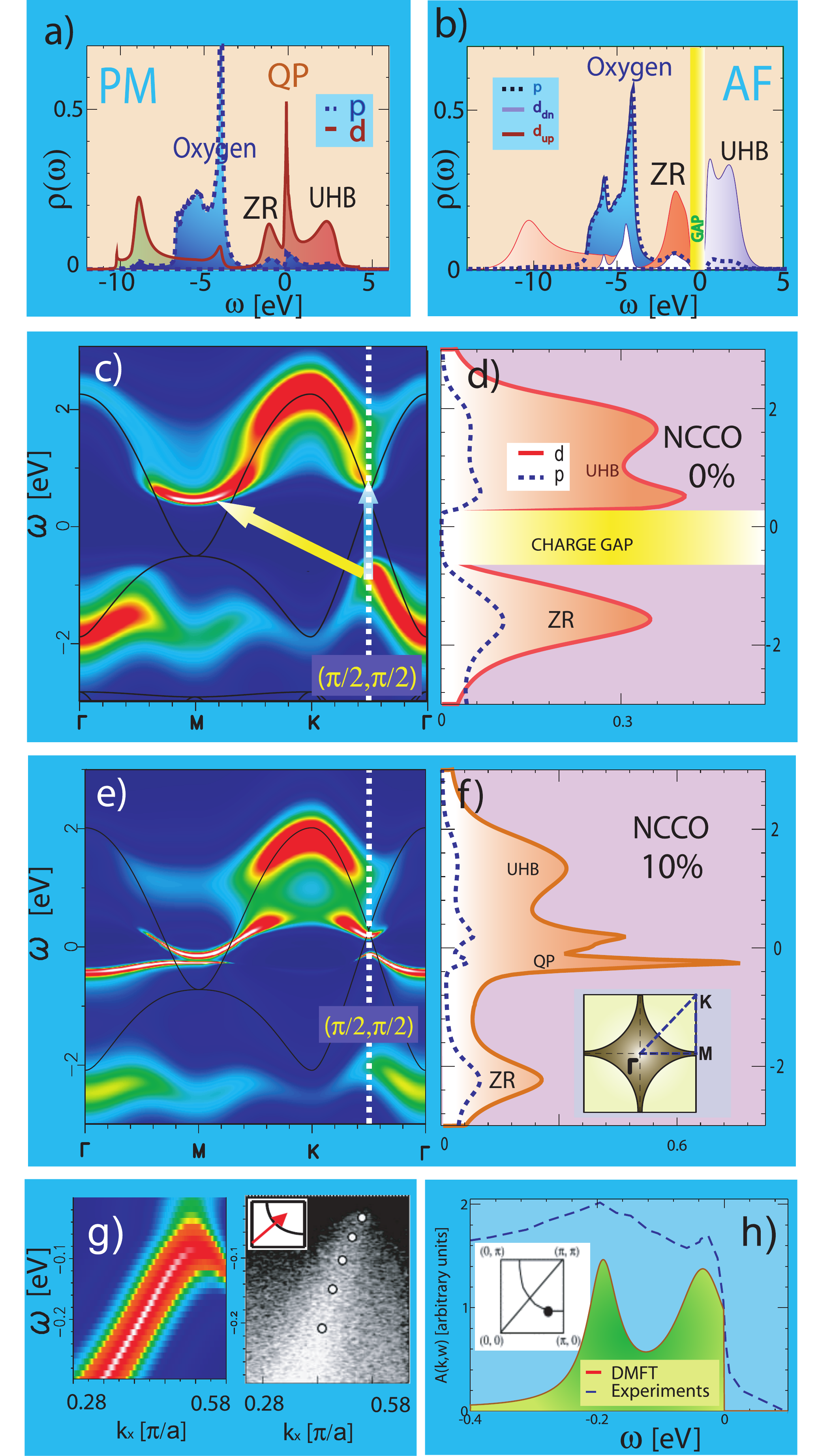}
\caption{ 
{\bf Density of electronic states in NCCO :}
(Colors online) Spectral function of NCCO a) nonmagnetic and b) magnetic) at zero doping in the extended
energy range. There is a gap of 1.2eV when magnetism is allowed. We clearly
observe the presence of the lower (upper) Hubbard bands LHB (UHB), 
and of a quasi-particle peak in the non-magnetic compound (QP).
Frequency dependant spectral weight A(k,w) along $\Gamma-M-K-\Gamma$
obtained by LDA+DMFT at c)
integer-filling and e) $10\%$ electron doping (NCCO). 
The partial density of
states of the d and p orbitals are shown on the right panels d) and f). 
The lines in c) and e) are the LDA
rigid bands, plotted in the folded Brillouin Zone, for comparison with DMFT.
The corresponding integrated weight is plotted in d) and f). 
g) side by side comparison of A(k,w), along the path $\Gamma-K$ as depicted in the inset, obtained
theoretically (left side) and experimentally from Ref.~\cite{arpes_comparison} (right side) at
$13\%$ doping.
h) comparison of A(w) for a fixed k point $k=(3\pi/4,\pi/4)$ (shown in the inset) 
obtained theoretically (lower curve) and experimentally from Ref.~\cite{arpes_comparison} (upper curve).
}
\label{fig:DOS}
\end{center}
\end{figure}
The results for the electronic spectral functions $A(\omega)=-1/\pi\textrm{Im}
G(\omega)$ of NCCO are shown in Fig.~\ref{fig:DOS}. The DMFT equations
have two solutions shown in panels \textbf{a,b}. The first one is
non-magnetic and the second is antiferromagnetically ordered. Notice
that the ordered solution, which is stable up to temperature $1400\,$K,
has a charge transfer gap of $1.2$eV.
When comparing to experiment, it is important to keep in mind
that two dimensional compounds are not able to sustain infinite-range
magnetic order at a finite temperature 
Therefore, the Neel temperature within DMFT
should be interpreted as the temperature below which the magnetic
correlations become long but remain finite. This temperature can be
much higher then the actual Neel temperature of the material, which is
controlled by the magnetic exchange between the two dimensional copper
oxide layers; and vanishes for a well separated copper oxide planes.
DMFT also allow us to investigate the underlying paramagnetic solution,
which describes a material in the absence of long range magnetic order.
This solution is metallic, and hence 
the magnetic long range order is responsible for the insulating nature of the compound,
as surmised by Slater~\cite{SOME_OLD_PAPER}.
It is worth noting that the oxygen orbitals carry no magnetic moment
because of the polarization with the two copper neighbors, which have
the opposite moments.

In Fig.~\ref{fig:DOS}\textbf{c} we resolve the low energy part of panel \textbf{b}
in momentum space. We find two dispersive coherent peaks separated by
the charge transfer gap. In Fig.~\ref{fig:DOS}\textbf{d} we resolve the low energy
part of panel \textbf{b} in orbital space. We find that the occupied
and empty states have mostly copper $d$ character with significant
oxygen $p$ admixture.
The top of the lower band occurs at $(\pi/2,\pi/2)$, while the bottom
of upper band apperars $M=(\pi,0)$, therefore the gap is indirect
(see yellow arrow in panel \textbf{c}).
Those two bands can also be obtained in the simpler Hartree Fock
approximation, though the size of the gap is overestimated in a static
mean-field.

In Figs.~\ref{fig:DOS}\textbf{e} and \textbf{f} we show the electronic structure
at $10\%$ electron doping. The compound is still magnetic. Some
aspects of the doped electronic structure can be understood in terms
of the Hartree Fock rigid band picture, the holes appear first in the
$M=(\pi,0)$ point, there are additional narrow quasiparticle states
close to but below the Fermi level. The two peak structure of the
low energy quasiparticles is clearly seen in the orbital resolved
angle integrated spectra of Fig.\ref{fig:DOS}~\textbf{f}. 

Notice also the presence of the 
pseudogap around $(\pi/2,\pi/2)$, that is a signature
of the presence of magnetic long range order, and 
is also observed in experiments \cite{arpes_comparison} (Fig.~\ref{fig:DOS}\textbf{g}, right panel),
and compares well to our theoretical calculations (Fig.~\ref{fig:DOS}\textbf{g}, left panel).
The pseudo-gap closes in the paramagnetic compound.
In Fig.~\ref{fig:DOS}\textbf{h}, we also show the spectral weight resolved in frequency for
a fixed $k=(3\pi/4,\pi/4)$ point (lower curve). 
We emphasize the ability of DMFT to resolve the 
multiple peak structure recently measured in experiments \cite{arpes_comparison} (Fig.~\ref{fig:DOS}\textbf{h} upper curve), 
that cannot be described in a mean-field picture like Hartree Fock.

\begin{figure}
\begin{center}
\includegraphics[width=0.8\columnwidth]{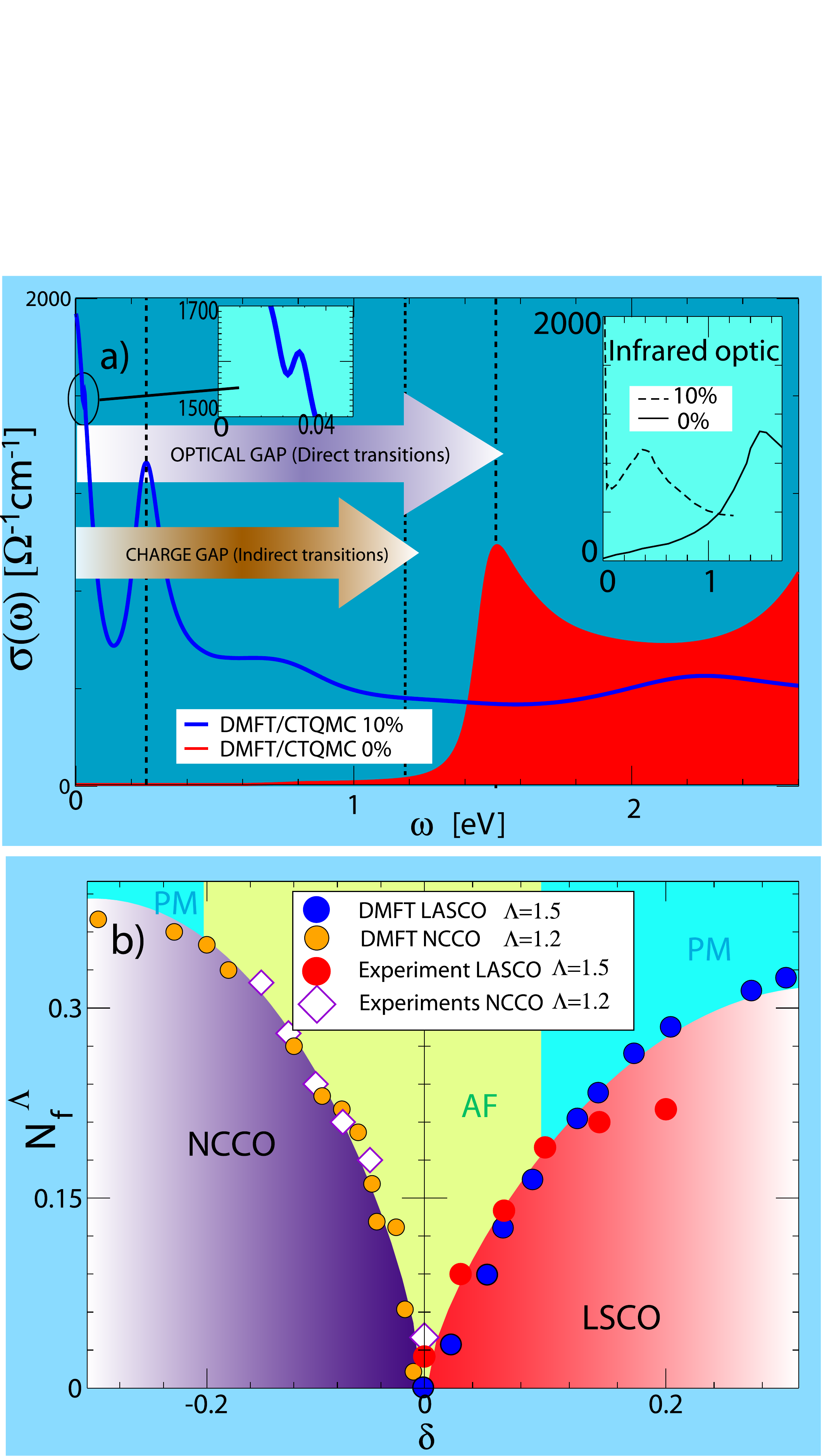}
\caption{
{\bf Optical properties of NCCO :}   
(Colors online)
a) Theoretical optical conductivity in $\Omega^{-1}cm^{-1}$ for NCCO doped with
charge carries concentration at integer filling (red curve) and at 10 percent doping (blue curve).
At integer-filling, the system is gapped and we observe an optical gap of about 1.5eV, that is larger than the charge gap $\approx 1.2$eV.
The left inset is a magnifying glass on the data.
For comparison we also show (right inset) the optical conductivity obtained by infrared optics \cite{infrared_optics}.
b) We show the dimensionless integrated optical conductivity $N_{eff}$ for LDA+DMFT done on LSCO and NCCO, obtained
with a cutoff $\Lambda=1.2$ ($\Lambda=1.5$) for NCCO (LSCO).
Experimental data for LSCO \cite{experimental_Nf} (red circles) and NCCO \cite{tokura_optics_ncco} (open diamonds) are shown. }
\label{fig:optic}
\end{center}
\end{figure}
We now turn to the theoretical optical conductivity, displayed in Fig \ref{fig:optic}\textbf{a},
which is in good qualitative agreement with the experimental results
reproduced in the right inset of Fig \ref{fig:optic}\textbf{a}.
The undoped compound has a sharp   onset at an energy
of the order of 1.5 eV
which we interpret as the direct gap (slightly larger than
the charge transfer gap in Fig.~\ref{fig:DOS}\textbf{d} and a tail resulting
from indirect transtions.
Doping introduces several new features. The $1.5 eV$  optical peak 
loses weight which is transferred to lower energy in the form of
a Drude peak and a mid infrared peak at  $\omega \approx 0.2eV$.

This peak can be explained in our theory by the presence of magnetism, that leads
to a reconstruction of the quasi-particle band structure as shown in Fig.~\ref{fig:DOS}{\bf e}).
in agreement with Ref.~\cite{spin_waves_ncco}, and additionally our study allow
to connect this peak with the spectral weight below the Fermi energy at the $M=(\pi,0)$ point.

Finally, the optical conductivity display a peak in the magnetic solution  
at a much lower  frequency $\omega \approx
0.03eV$ (see left inset of Fig.~\ref{fig:optic}\textbf{a}), 
which is connected to the pseudo-gap at $\bold
k=(\pi/2,\pi/2)$, and is also observed in experiments
\cite{tokura_optics_ncco}.

To quantify the rate of the redistribution of optical spectral weight,
is useful to consider the effective electron number per Cu atom
defined by $ N^{\Lambda}_{eff}=\frac{2 m_e V}{\hbar \pi e^2}
\int^{\Lambda}_{0}{\sigma'(\omega) d\omega}, \label{Sweight}$ where
$m_e$ is the free electorn mass, and $V$ is the cell Volume containing
one formula unit. $N_{eff}$ is proportional to the number of electrons
involved in the optical excitations up to the cutoff $\Lambda$.  Our
results for (electron doped) NCCO are displayed in the left hand side
of Fig.~\ref{fig:optic}\textbf{b} and are compared to experimental data taken from
Ref.~\cite{tokura_optics_ncco}.
Notice a favorable agreement between the theory and experiment, for
which the use of the realistic electronic structure is essential.
In Fig.\ref{fig:optic}\textbf{b} (right) we also show our theoretical optical spectral weight
for a prototypical hole doped cuprate (LSCO), 
computed in Ref.~\cite{our_previous_paper_lsco}, which is a Mott type
insulator, and we compare it to experimental
data~\cite{experimental_Nf}. The results for a Slater type insulator
NCCO and Mott type insulator LSCO are very similar, hence we can
conclude that the optical spectral weight is not a precise measure of
the strength of correlations.

\begin{figure}
\begin{center}
\includegraphics[width=0.8\columnwidth]{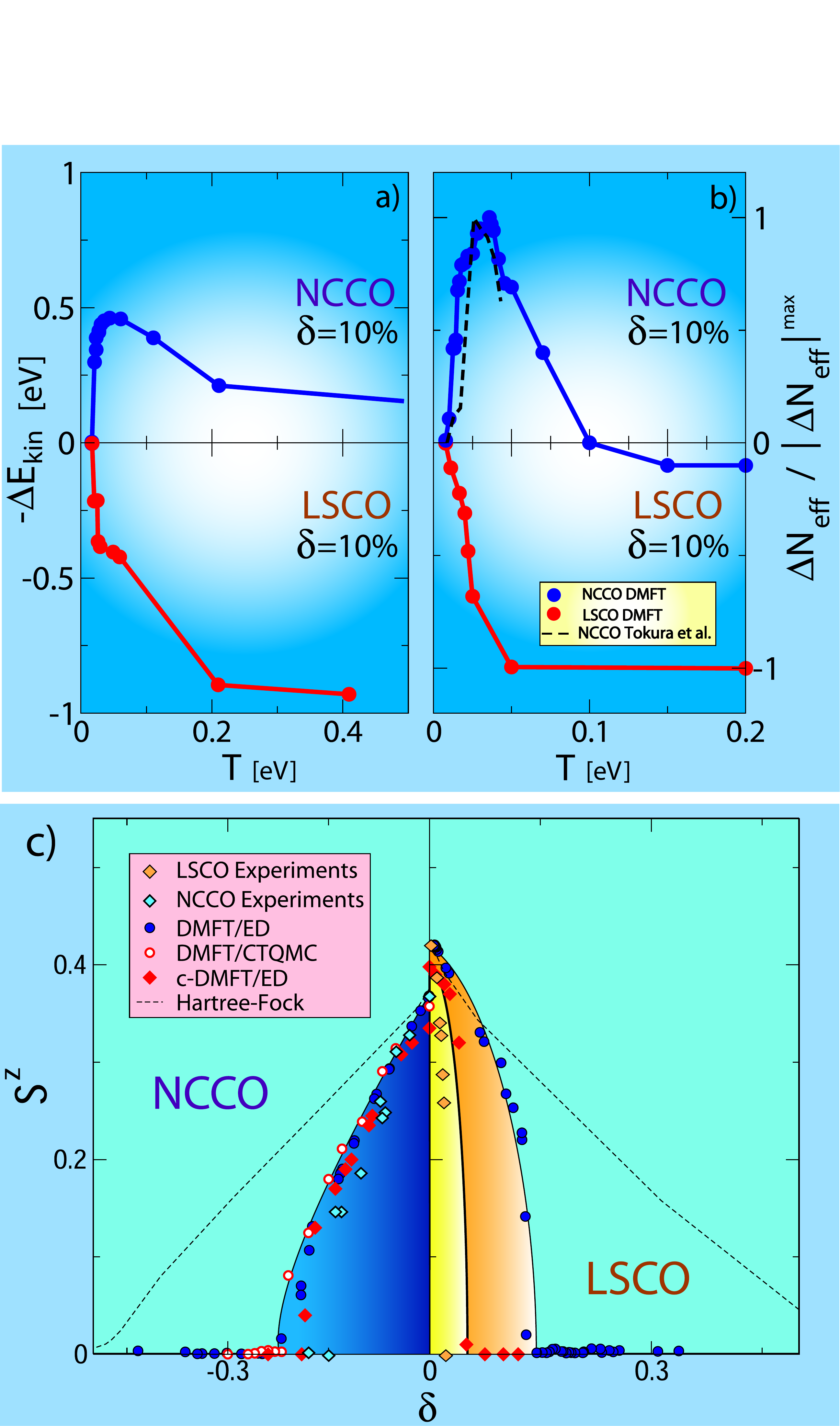}
\caption{
{\bf Kinetic energy and magnetic phase diagram: } 
(Colors online)
a) Theoretical kinetic energy variation $-\Delta E_{kin} = -E_{kin}(T)+E_{kin}(89^\circ K)$ of 
the electron (top panel) and the hole (bottom panel) doped compounds with
respect to temperature.  
b) Normalized variation of $N_{eff}$, $\Delta N_{eff} = N_{eff}(T)-N_{eff}(89^\circ K)$
of the electron (top panel) and the hole (bottom panel) doped compounds with
respect to temperature. The dashed line are extracted from experiments done
on NCCO \cite{tokura_optics_ncco}.
c) Mean-value of the staggered magnetization obtained for LSCO and NCCO by
both single site (DMFT) and cluster DMFT (c-DMFT). 
Experimental values $M(\delta)/M_0$ ($M_0=M(\delta=0)$) for NCCO \cite{staggered_moment} and LSCO \cite{lsco_magnetic_moment} are also shown,
and for comparison with DMFT, we assume $M_0=M_{DMFT}(\delta=0)$, where $M_{DMFT}$ is the magnetic moment at 0 doping obtained
by single site DMFT.
We also show the magnetization obtained in the
static hartree-fock approximation (dashed line) are shown for comparison.
}
\label{phasediag}
\end{center}
\end{figure}

In a Slater picture, the onset of antiferromagnetism reduces the
Coulomb correlations (double occupancy) at the expense of the kinetic
energy.  The opposite is true in a Mott insulator. Consequently, in a
Slater insulator the kinetic energy becomes less negative as the
temperature decreases while the opposite happens in a Mott insulator.
The kinetic energy as a function of temperature is readily available
in the theory and is displayed in Fig~\ref{phasediag}\textbf{a} (top panel).

A closely related quantity $N_{eff}$ is experimentally accessible and
has been measured for this compound. In Fig~\ref{phasediag}\textbf{b} (top panel) 
we plot both the experimental ~\cite{tokura_optics_ncco} and our theretical data, which are in very
good agreement. 
Experiments of Ref.~\cite{tokura_optics_ncco} 
confirm that $N_{eff} $ is closely related to minus the
kinetic energy. It has therefore already been confirmed that NCCO is below
the metal to charge transfer insulator transition in experiments.

In Fig.~\ref{phasediag}\textbf{a} (bottom panel)
we also plot the temperature dependence of the kinetic
energy and $N_{eff}$ for the hole doped Mott insulator (LSCO). We
find the opposite trend compared to the lightly electron doped NCCO
compound, confirming that LSCO is a doped Mott insulator.

Finally, in order to check the validity of our single site DMFT
approach, we computed the magnetic phase diagram and magnetic moment
within the single site and two-site cluster DMFT (see
Fig.~\ref{phasediag}.c).  Our data for NCCO are displayed on the left
side of Fig.~\ref{phasediag}.c) and are compared to experimental data from
Ref.\onlinecite{staggered_moment}. The agreement is very good for both approaches,
hence justifying the use of the single sited DMFT approach.

The right site of Fig.~\ref{phasediag}.c) displays theoretical and experimental
magnetic moment\cite{lsco_magnetic_moment} for hole doped LSCO. We found significant difference
in the region of stability of the magnetic state between the single
site and cluster DMFT ($\delta<10\%$) , hence the dynamical short
range correlations - absent in single site approach - are very
important for LSCO in the underdoped regime, but not for NCCO.

We also carried out the Hartree-Fock calculation, and we found that in
this static approach the magnetization vanishes only at unrealistic
large doping $\delta\approx50\%$ for both NCCO and LSCO (dashed line
in  Fig.~\ref{phasediag}.c), which points towards the important role of dynamic
correlations at finite electron and hole doping.

Our theory sheds light on many puzzling observations. 
It was noticed within the context of the one band model that the
doping dependent Hubbard interaction was needed to reproduce
experiments~\cite{kyung_U_dependant}. In our view, this is a
description, within a one band model, of the rapid metallization
process in a Slater-like insulator. A realistic LDA+DMFT treatment of
the multiband Hamiltonian does not require doping dependent
interactions.
The strong sensitivity of the electronic
properties to magnetism, which results from the 
fact that the physics of NCCO is closer to that
of a Slater insulator than of a Mott insulator, gives rise to additional
experimental predictions. We expect that at high temperatures
(of the order of the DFMT mean field temperatures) there should
be appreciable transfer of spectral weight from high energies to
low  energies with decreasing temperature,
as the magnetic correlations are weakend,
and a substantial decrease of the Mott Hubbard gap
and the weights in the Hubbard bands.
This can be checked by extending transport
and photoemission  studies
in lighlty doped electron doped cuprates from 600 to 1100 K.

In materials where the strength of the interaction is below
the critical value need to produce a Mott insulating state,
the effective interactions can renormalize down, allowing
superconductivity and magnetism to coexist microscopically rather
than exclude each other. This coexistance provides a good description
of the Raman scattering of the electron doped superconductors ~\cite{raman_electron_doped}. 
Finally, in some NCCO films the
parent compound was found to be metallic. It would be interesting
to study the magnetic  properties of these materials to see if this metallicity
correlates with a substantial decrease of the magnetic correlations ~\cite{Blumberg}.

We thank A.M. Tremblay, D. Basov, D. G. Hawthorn, G. A. Sawatzky
for discussions and sharing their insights and experimental data. Numerous
discussions with A. Georges,  A. Amaricci, J. C. Domenge and A.
Millis are gratefully acknowledged. Adriano Amaricci shared his
density matrix renormalization group code and Jean-Christophe
Dommenge shared his exact diagonalization code. K.H was
supported by Grant NSF NFS DMR-0746395 and Alfred P. Sloan
fellowship. G.K. was supported by NSF DMR-0906943, and C.W. was
supported by the Swiss National Foundation for Science (SNFS).

\section{Method}

The LDA calculation was carried out with the PWSCF
package~\cite{cappuccion}, which employs a plane-wave basis set and
ultrasoft pseudopotentials~\cite{lda_vanderbilt}. We downfold to a
three band model, containing copper \dx and two oxygen \psig orbitals
using the maximally localized Wannier functions method
(MLWF)~\cite{lda_basis}. The results of this procedure can
be summarized in the following three band Hamiltonian:
\begin{multline}
\label{eq:3band_hub}
\mathcal{H}= \sum\limits_{ij\sigma, (\alpha,\beta) \in (p_x,p_y,d_{x2-y2})}{  t^{\alpha \beta}_{ij} c^\dagger_{i \alpha \sigma} c_{j \beta \sigma} }
          + \epsilon_p \hat N_{p} + \\
 \left(\epsilon_d-E^{dc}\right) \hat{N}_{d} + U_d \sum_{i}{ \hat n_{id\up} \hat n_{id\dn}}
\end{multline}
where $i$ and $j$ label the CuO$_2$ unit cells of the lattice, and
$t_{ij}^{\alpha\beta}$ are the hopping matrix elements, and
$\epsilon_d$, $\epsilon_p$ are the on-site energies of the $d$ and $p$ orbitals.
The onsite Coulomb repulsion $U$ on the \dx orbital was fixed to
$U_d=8 eV$. Finally, we note that $\epsilon_d-\epsilon_p$ plays the role of an
effective onsite repulsion $U$ in a Hubbard model picture.
The LDA+DMFT method, accounts for the correlations which are included
in both LDA and DMFT by a double counting correction to the
$d$-orbital energy, $E_{dc}$ which amounts to a shift of the relative
positions of the d and p levels which we take to be doping independent
and $E_{dc}=4.8eV$ ($3.12eV$) for NCCO (LSCO). The LDA downfolded
hopping parameters are presented in the online material.
The Green function of the three band model is given by:
\begin{equation}
 \label{greenfunc}
 \textbf{G}_\vk(i \omega_n) =  \left( i \omega_n + \mu - \bold{H}_\vk - \bold{\Sigma}(i \omega_n)   \right)^{-1},
\end{equation}
where $\bold{H}_\vk$ is the Fourier transform of the $\mathcal{H}_t$
in Eq.~(\ref{eq:3band_hub}) and $\bold{\Sigma}$ is the DMFT
self-energy matrix, which is assumed to be local, and nonzero only for
the $d$ orbital.
The self energy in Eq.~(\ref{greenfunc}) is obtained by solving an
Anderson impurity model subject to the DMFT self-consistency
condition
\begin{equation}
 \frac{1}{ i\omega - E_{imp} - \Sigma(i\omega) - \Delta(i\omega)}  = \frac{1}{N_k} \sum\limits_{\vk \in BZ}  {G_\vk^{dd}(i \omega)},
\end{equation}
where the sum runs on the first Brillouin Zone (BZ). The self-energy
$\bold{\Sigma}$ is obtained by solving an Anderson impurity using the
continuous time quantum Monte Carlo impurity
solver~\cite{werner_ctqmc_algorithm,Haule_long_paper_CTQMC} (CTQMC) and the exact
diagonalization solver~\cite{OLD_GABIS_REVIEW} (ED). Calculations have been
carried out at temperature $T=89^\circ K$, when not specified in the
text.
Finally, the optical conductivity is given by
\begin{multline}
 \sigma'(\omega) = \frac{1}{N_k} \sum\limits_{\sigma \vk}{\frac{\pi e^2}{\hbar c }  \int{ dx \frac{  f(x-\omega)-f(x)}{\omega}}} \\
                   \times \textrm{Tr}\big( \hat{\bold{\rho}}_{\vk\sigma} (x - \omega)  \bold{v}_\vk  \hat{\bold{\rho}}_{\vk\sigma}(x) \bold{v}_\vk \big),
\end{multline}
where $c$ is the interlayer distance, the density matrix $\bold{\hat{\rho}}$ is 
$ \hat{\bold{\rho}}_{\vk\sigma}(x)=\frac{1}{2\pi i} \left(\bold{G}^\dagger_{\vk\sigma}(x)-\bold{G}_{\vk\sigma}(x)\right) $,
and the bare vertex is obtained by following the steps of Ref \cite{ian}.


\bibliographystyle{prsty}

\begin{thebibliography}{10}

\bibitem{armitage_review}
N.P. Armitage et al., Phys. Rev. B. {\bf 68}, 064517 (2003).

\bibitem{takagi}
H. Takagi et al., Phys. Rev. Lett. {bf 62}, 1197(1989).

\bibitem{tokura_electron_doped}
Y.~Tokura and A.~Fujimori, Phys. Rev. B {\bf 39}, 9704 (1989)

\bibitem{our_previous_paper_lsco}
C. Weber, K. Haule, and G. Kotliar , Phys. Rev. B. {\bf 78}, 134519 (2008)

\bibitem{OLD_GABIS_REVIEW}
A. Georges et al., Rev. Mod. Phys., {\bf 68}, 0034 (1996)

\bibitem{lda_dmft_held}
K Held et al.,  J. Phys.: Condens. Matter, {\bf 20}, 064202 (2008)

\bibitem{arpes_comparison}
H. Matsui et al., Journal of Physics and Chemistry of Solids, {\bf 67} 249 (2006).

\bibitem{SOME_OLD_PAPER}
J. C. Slater, {\it Quantum theory of molecules and solids}, (McGraw-Hill, New-York, 1974), Vol. 4.

\bibitem{infrared_optics}
T. Xiang et al., Phys. Rev. B {\bf 79}, 014524 (2009).

\bibitem{experimental_Nf}
S. Uchida {\it et~al.}, Phys. Rev. B {\bf 43},  7942  (1991).

\bibitem{tokura_optics_ncco}
  Y. Onose et al., Phys. Rev. B, {\bf 69}, 024504 (2004).

\bibitem{spin_waves_ncco}
 A. Zimmers et al, Europhys. Lett. 70, pp. 225-231 (2005).

\bibitem{staggered_moment}
P. K. Mang {\it et~al.}, Phys. Rev. Lett. {\bf 93}, 027002 (2004).

\bibitem{lsco_magnetic_moment}
F. Borsa et al., Phys. Rev. B. {\bf 52}, 7334 (1995).

\bibitem{kyung_U_dependant}
B. Kyung et al., Phys. Rev. Lett. {\bf 93}, 147004 (2004)

\bibitem{metal_ncco_experiments}
O. Matsumoto et al., Phys. Rev. B, {\bf 79}, 100508 (2009)

\bibitem{ncco_swave}
Guo-meng Zhao, condmat/0907.2011

\bibitem{yung}
Y. Qingshan et al.,  Phys. Rev. B {\bf 72}, 054504 (2005)

\bibitem{raman_electron_doped}
B. Stadlober et al., Phys. Rev. Lett. {\bf 74}, 4911 (1995).

\bibitem{Blumberg} 
 J.-P. Ismer et al. , arXiv: 0907.1296.

\bibitem{cappuccion}
PWSCF in Quantum Espresso Package (2007) URL http://www.pwscf.org/  .

\bibitem{lda_vanderbilt}
D. Vanderbilt, Phys. Rev. B {\bf 41},  7892  (1990).

\bibitem{lda_basis}
I. Souza, N. Marzari, and D. Vanderbilt, Phys. Rev. B {\bf 65},  035109 (2001).

\bibitem{Haule_long_paper_CTQMC}
K. Haule and G. Kotliar, Phys. Rev. B {\bf 76},  104509  (2007).

\bibitem{werner_ctqmc_algorithm}
P. Werner {\it et~al.}, Phys. Rev. Lett. {\bf 97},  076405  (2006).

\bibitem{ian}
Jan M. Tomczak, Eur. Phys. Lett. {\bf 86}, 37004 (2009).

\end{thebibliography}

\end{document}